# On-Line Selection and Quasi-On-Line Analysis of Data using the Pixel Lensing Technique

S. Capozziello[*] and G. Iovane[+]

Dipartimento di Scienze Fisiche "E.R.Caianiello"

Università di Salerno, I-84081 Baronissi, Salerno, Italy

INFN Napoli, Gruppo Collegato di Salerno, Italy

**Abstract**

Pixel lensing is a technique for searching baryonic components of dark matter, (MACHOs), in any part of galaxies. This method is very efficient, since it avoids to resolve each source star and so it has the advantage to provide an high statistic. Taking into account the fact that this technique is very time expensive in data analysis phase and the present technology, we propose a new method to perform a data selection on-line by using two levels of trigger and a quasi-on-line data analysis. The implementation of the project provides to obtain soon results since, due to a new database technology, all physical information are stored for a complementary and secondary analysis to perform off-line.

---

[*] E-Mail: Capozziello@vaxsa.csied.unisa.it
[+] E-Mail: Geriov@vaxsa.csied.unisa.it

# 1 Introduction and Present Scenario

In recent years, much attention was centred on the possibility that dark matter, and in particular its baryonic side, consists of astrophysical objects, generically termed as "Massive Astrophysical Compact Halo Objects" (MACHOs) with mass $10^{-8} M < M < 10^2 M$. In this large mass range, they could be primordial black holes, brown dwarfs, remnants of "Population III" stars (e.g. white dwarfs), Jupiters or Snowballs.

Direct searches of these objects can, at best, reach the solar neighbourhood. In order to detect them further out, it was proposed by Paczynsky (1986) to search for dark objects by gravitational lensing [1]. In fact, when a compact object passes near the line of sight of a background star, the luminosity of this star increases giving rise to a characteristic luminosity curve. The deflection of the light ray is given by

$$\alpha = \frac{4GM}{c^2 R} = \frac{2R_s}{R}, \qquad (1)$$

where $R_s$ is the deflector Schwarzschild radius and $R$ is the distance between the deflector and the light ray, i.e. the impact parameter. In other words, the amplification $A$ is

$$A = \frac{R_E}{x}, \qquad (2)$$

where $R_E = \sqrt{\frac{2R_s d_{LO} d_{LS}}{d_{LO} + d_{LS}}}$ is the Einstein radius with $d_{LO}$ the distance between the lens and the observer, $d_{LS}$ the distance between the lens and the source and $x$ is the impact parameter [2].

From this point of view the gravitational lensing is a direct implication of general relativity. In particular, we deal with microlensing when the separation among the images is too small to be detected. We have separations $\delta\vartheta \leq 10^{-3}$ arcsec for galactic microlensing and $\delta\vartheta \leq 10^{-6}$ arcsec for extragalactic microlensing. This justifies the name.

It is important to stress that the detection of a high number of microlensing events is one of the best way to investigate the Galactic structure and it would certainly allow to determine whether MACHOs are clustered or diffused in the Galactic halo.

In this context, an important goal is to perform a map of distribution of MACHOs in the galactic bulge, disk and halo of the Milky Way and furthermore in and, M31, M33 and dwarf galaxies in the Local Group.

Until now more than one hundred microlensing events have been detected by some collaborations.

EROS (Espérience de Recherche d'Objets Sombres)[3] and MACHO (Massive Astrophysical Compact Halo Objects)[4] are looking toward LMC and SMC, DUO (Disk Unseen Objects)[5] and OGLE (Optical Gravitational Lensing Experiment)[6] and MACHO too are monitoring the bulge of the Galaxy, AGAPE (Andromeda Galaxy Amplified Experiment)[7] and Columbia-VATT (Columbia-Vatican)[8] are looking toward M31.

The last two experiments are particularly relevant since they are using the new detection technique called "Pixel Lensing", which we describe in sections below. Essentially it allows to search for microlensing events considering the pixels of a CCD camera without taking into account the problem of resolving the source stars. In this way, also very crowded fields, like the bulge of Andromeda galaxy, can be considered for microlensing issues.

Despite of this huge advantage, the actual researches have not given a great deal of useful results since the data analysis is cumbersome and takes a lot of time .

In this paper, we propose a new technique for an on-line selection and a quasi-on-line analysis in order to improve the pixel lensing technique . The proposal is essentially direct to the future experiment like SLOTT-AGAPE (Systematic Lensing Observations Toppo Telescope - Andromeda Galaxy Amplified Experiment) which is going on to run at the Telescope of Napoli Observatory in Castel Grande (Potenza) in the South of Italy.

The scientific goals are several (e.g. microlensing events, planet detections, supernovae searching, etc.) and most of them should be use improved versions of pixel lensing technique.

The paper is organised as follows:

- In Sec.2, we deal with the conceptual design of data acquisition and handling;
- Sec.3 is devoted to the standard pixel lensing technique;
- In Sec.4, we find an instrumental hardware note;
- Sec.5 is devoted to the preliminary project for the data acquisition, selection and analysis;
- At the end, we find the section: conclusion and perspective.

## 2 Conceptual Design: Data acquisition and Handling

We propose to implement a new project for data acquisition, selection and handling. A two level trigger follows the data acquisition to perform the on-line selection of interesting events that can be studied quasi-on-line. Other events, which should be more complicate lensing events, variable stars, novae and supernovae and so on, are not lost during the selection, but are stored for an off-line analysis[1].

Therefore we would like to stress that, besides the advantages of an on-line selection and quasi-on-line data analysis, the possibility to perform the analysis of other interesting events is saved.

The idea is to use a technique which is an evolution of the pixel lensing implemented by AGAPE starting from 1992. The previous method is considered inside a more general environment, where performing the analysis on the single pixel of the CCD camera, a multi-fit system is implemented instead of the simple Paczynski fit. In this way, we can take into account different lensing events as binary lenses or binary sources, in the quasi-on-line analysis and the possibility to have more involved lensing events or other events with an increase of the light curves.

Such a project requires the implementation of new software tools, as a "relational database" [9] or better an "object database" [10], where it is possible to store the data and prepare next observations. Since such a database has to be connected with a dispatcher, it could be possible to implement an automatic system to alert the operator about an interesting event that is obtained after the standard operation of geometric and photometric calibration, which we will illustrate below.

From this point of view the database is the main software element for the data acquisition, processing and handling becoming the connection between the telescope, the CCD camera, the hardware control (Data Acquisition Unit: DAQ ) and the software for the event selection and analysis (Process and Analysis Unit: P&A).

In other words, our system should be a semiautomatic unit which could perform a selection and a pre-analysis without any attendance operation. In this way, the researcher should study and certify only really interesting events, obtaining soon after the data acquisition some interesting physical results[2].

---

[1] Standard pixel lensing technique, till now, is not considering such kinds of events since it is not able to handle them.

As we shall see in the next pages these are not the only advantages of our project.

**3 The Standard Pixel Lensing**

Searching for microlensing requires to monitor a huge number of stars. The idea of pixel lensing originated in order to carry out this request using M31 as a target star field.
Two approaches have been develop independently:
- the monitoring of pixel light curves implemented by the AGAPE collaboration;
- the image subtraction techniques implemented by Columbia/VATT collaboration.

In this paper we will refer to the monitoring of pixel light curves.

In a dense field, many stars contribute to any pixel of the CCD camera at the focal point of the telescope. When an unresolved star is sufficiently magnified, the increase of the light flux can be measured on the pixel. Therefore, instead of monitoring individual stars, we follow the luminous intensity of the pixels. Then all stars in the field, and not the only few resolved ones, are candidates for a microlensing event; so the event rate is potentially much larger. Of course, only the brightest stars will be amplified enough to become detectable above the fluctuations of the background, unless the amplification is very high and this occurs very seldom. In a galaxy like M31, however, this is compensated by the very high density of stars.

The main difficulties to overcome are the variations in observing conditions:
- On two successive images a pixel never points toward the same part of the sky. In this case, it is important to make a geometrical alignment. The transformation is roughly a translation plus a rotation,

$$x' = Ax + \Delta x,  \qquad (3)$$

  where, generally, the matrix $A = \begin{pmatrix} a & b \\ c & d \end{pmatrix}$ contains possible shear effect or other effect due to the electronic device besides the rotation. For such an alignment we can use the stars in front of M31.

---

[2] At this moment the researchers working at TNG (Telescopio Nazionale Galileo) are considering the possibility to implement the on-line alignment; taking into account the physical requirement we would like to realise for SLOTT-AGAPE [11] the previous proposal studies about the selection on-line and the analysis quasi-on-line in addition to the calibration.

- Two frames are never taken in the same photometric conditions due to the atmospheric absorption or night sky luminosity.

In first approximation, photometric alignment is performed assuming that all differences in instrumental absorption between runs have been removed by the correction for flat fields. In this case, one may assume the existence of a linear relation between the intensity $F$ in corresponding pixel of the current and reference images:

$$F_{pixel}^{reference\ image} = aF_{pixel}^{current\ image} + b, \qquad (4)$$

where **a** is the ratio of absorptions and **b** the difference of sky backgrounds between the reference and the current image.

The usual way to evaluate **a** is to compare the total intensities of the corresponding stars on the two pictures.

- After the two alignments above, an intensity gradient remains between successive images which can be deleted using an equalizing process on the images.

- Finally, there are some changes in the seeing between successive images. Seeing variations induce fake variations, therefore must be corrected for as much as possible.

The first possible correction is to use a large super-pixel to compare to the average value of the seeing, so that most of the star flux should be included in one super-pixel, but not too large to avoid diluting the signal. Using 2.1''×2.1'' superpixels turned out as the best trade-off. This assumption corrects part of the variations, but it is now able to do much better using the "Paris Method" or "Geneva Method" (see AGAPE analysis reference). Here we will describe the Paris seeing correction. This correction relies on an empirical observation: the difference between the values of the flux of the same pixel on the current image and on the reference image, $\left(\phi_{pixel}^{current} - \phi_{pixel}^{reference}\right)$, is linearly correlated to the same difference, but between the reference and a median[3] of the current image, $\phi_{pixel}^{reference} - \phi_{pixel}^{reference}$ :

$$\phi_{pixel}^{current} - \phi_{pixel}^{reference} = \alpha\left(\phi_{pixel}^{reference} - \phi_{pixel}^{median}\right). \qquad (5)$$

The median behaves as if it separates regions with different behaviours with respect to the seeing. In particular, when the reference image is brighter than the median image, then good seeing makes the current image even brighter whereas bad seeing makes it fainter, but good and bad seeing exchange each other when the reference image is darker than the median. The correlation and the variation of the slope $\alpha$ with seeing can be understood as follows: when the seeing becomes worse, the image becomes smoother and nearer to its median. If one assumes that the difference between any image and its median decreases like some inverse power of the seeing:

$$\phi_{pixel}^{image} - \phi_{pixel}^{median} \propto \frac{1}{\left(seeing^{image}\right)^{\eta}} \tag{6}$$

one gets:

$$\alpha = \left(\frac{seeing^{reference}}{seeing^{image}}\right)^{\eta} - 1. \tag{7}$$

---

[3] The median of an image is obtained by replacing each pixel by the median of the flux of the pixel inside a $41\times 41$ window centered on the original pixel.

In order to implement the correction for each image, $\alpha$ is determined from a linear fit while the reference image is replaced by the corrected image:

$$\phi_{pixel}^{corrected} = \frac{\phi_{pixel}^{current} + \alpha\, \phi_{pixel}^{median}}{\alpha + 1}. \tag{8}$$

The first step for the analysis of light curves is to define the baseline on the background flux $\phi_{BKG}$. This is done by taking the minimum of a sliding average on 10 consecutive points. One can, then define a bump as beginning when 3 consecutive points lie $3\sigma$ above $\phi_{BKG}$ and ending when 2 consecutive points fall below $3\sigma$. One can define a likelihood function $L_{bump}$ for each bump:

$$L_{bump} = \ln\left(\prod_{j \in bump} \rho(\phi|\phi\rangle\phi_j)\right) \quad \text{given} \begin{Bmatrix} \phi_{BKG} \\ \sigma_j \end{Bmatrix}. \tag{9}$$

The second step is to select microlensing candidates by light curves with only one bump: this is defined as a light curve with at least one bump having $L_1 > 300$ and no second bump with $L_2 > 70$.

The third step is to fit a high amplification degenerate Paczynski curve to the mono-bumps. The amplification is then well approximated by

$$A(t) - 1 \simeq \frac{1}{u(t)} \quad \text{with} \quad u(t) = \sqrt{\left(\frac{t - t_0}{t_E}\right)^2 + u_0^2}, \tag{10}$$

where the Einstein time is $t_E = R_E / V_T$, the ratio of the Einstein radius $R_E$ to the transverse velocity of the lens $V_T$. The formula for the flux of the super-pixel is:

$$\phi_{spx} = \phi_{BKG} + \frac{\phi_*(A_{max} - 1)}{\sqrt{\left(\frac{t - t_0}{t_E / (A_{max} - 1)}\right)^2 + 1}} \quad , \tag{11}$$

where $t_0$ is the instant of maximum amplification $A_{max}$, and $\phi_*$ is the flux of the amplified star "at rest". Actually $A_{max}$ and $t_0$ are difficult to be exactly determined for a pixel event. Eq. (11) holds as well as the Paczynski fit is good. The results can be improved if one define a set of several likelihood function $L^i_{bump}$ with $i = 1, ..., N$ able to perform several fits.

## 4 An instrumental hardware note

At this point it could be clear what are the conditions to perform an efficient microlensing survey using M31. In particular:

a) the field of the instrument should be large enough to be able to cover the whole useful part of the galaxy with a small number of exposures. This useful field is roughly 50'×15'.
b) The average seeing should be equal or better than 1.5", since a large seeing dilutes the light of the amplified star. The stability of this seeing is also important.
c) The pixel size should be small compared to the average seeing, to allow geometrical alignment without degrading the image quality. Pixels 0.4" wide will probably be good for a typical seeing of 1.5".
d) Sampling time is an important point too. It should be every one or two nights for the period when M31 is high enough in the night sky.
e) At the end, in our opinion, a period of three consecutive years should be enough to obtain a good statistics.

The use of the Toppo telescope $(D = 1537\,mm, \ F/8.5)$ with an available square field of view of 24'×24' (corresponding to a 30' diameter) appears particularly well adapted for the pixel lensing technique since the inner regions of M31 are imaged in the 12'×12' field of view of the available CCD camera. Furthermore for this kind of survey, where it is important the achromaticy test, a beam splitter is welcomed to perform simultaneous observations in the red and blue bands.

# 5 The Preliminary Project for Data Acquisition, Selection and Analysis

In our project the system is made by different units as shown in the following scheme.

In such a system we have:

i) The Data Acquisition (DAQ) Unit that is responsible of the data acquisition.

ii) The Control Unit, which thanks to Telescope Control System (TCS), controls the telescope by following the instruction from the DataBase Control System (DBCS) or console.

iii) The DataBase (DB) Unit is the intelligent part of the system, in fact, here we have the data storage and processing. Such a unit is the natural link from the observations and the data analysis.

iv) The Processing and Analysing (P&A) Unit is the platform where the massive data analysis is made.

The dotted part (e.g. Telescope Remote Control, Telescope Remote Observing) in the control unit could means too futuristic; actually it could be very useful if one starts to think about remote control and observations.

From a conceptual point of view the schematic steps of the project of data flux are the following:

**a) Observation, measurement and data acquisition:**

a.1) Seeing and meteo monitoring in real time.

a.2) Acquisition of reference images:

        i) bias

        ii) dark

        iii) flat - field

a.3) Acquisition of the real image corresponding to the star field of our study.

**b) Image Processing and filtering.** This operation is made for any pixel of the CCD camera.

$$I^{ij}_{\text{effective}} = I^{ij}_{\text{scientific}} - I^{ij}_{\text{technical}} \quad \text{with} \quad i, j = 0, \ldots 2048 \,, \quad (12)$$

where $i$ and $j$ are the row and column index of such pixel.

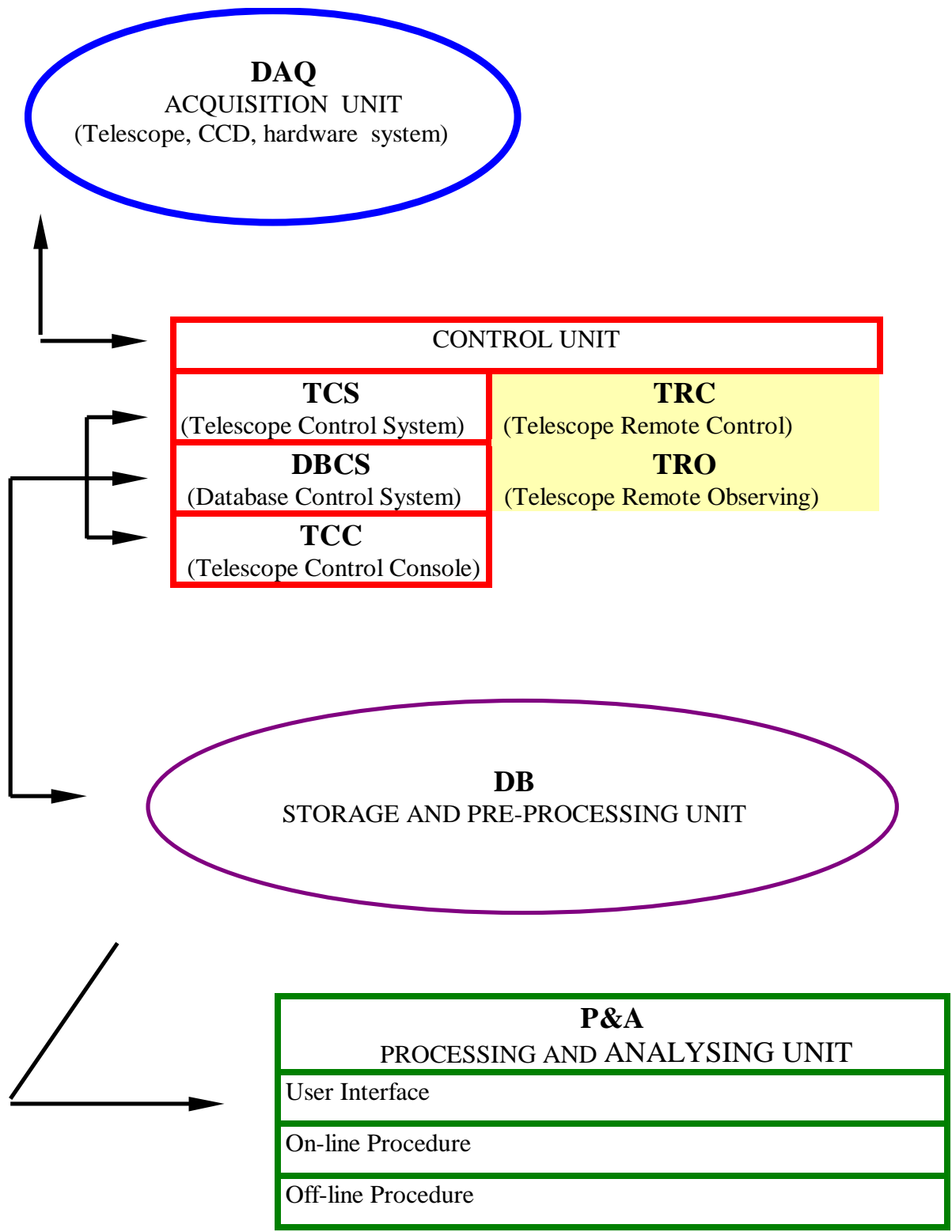

**The scheme of the project.**

About this operation, we believe that it could be applied a pipeline procedure; in such a case, we will need an analysis to understand the quality of data reduction. It can be done with an automatic system using a known calibrator or an Artificial Neural Network (ANN). This set of operations is named technical calibration; in this step it is removed the hardware noise (electronic):

- pixel defects;
- quantum inefficiency;
- non linear effect in the output.

Concerning the output calibration, we have:

$$X^{ij}_{tecnica} = \frac{C^{ij} - D^{ij}}{F^{ij} - D^{ij}} \quad , \tag{13}$$

where C is the raw image, D is connected with the dark counting and F is the flat-field exposition. Therefore we can look at X as an image of relative intensity. The output of an element of the CCD camera will be:

$$I^{ij}_{tecnica} = M^{ij} \times X^{ij} + A^{ij} + F^{ij} \times X^{ij} \quad , \tag{14}$$

where X is the input to this step, M is a multipling coefficient that take into account the quantum efficiency, A is an adaptive term for the dark and F contains the non linear effects.

**c) Corrected raw data saving in the database.**

**d) Evaluation of all quantities phisically interesting.**

After the sequence of such operations is made for the first observation, the database prepares the expected conditions for each pixel in the next observations; in other words, it has to prepare the calibration useful to detect interesting events.

**e), f), g), h)** are like a), b), c), d) in the observation that follow the first one.

**i) Geometrical calibration.** The geometrical alignment is obtained as written above.

**l) Photometric calibration.** Also in this step we go head as before shown.

**m) First Level Trigger** (Selection of Interesting Events).

m.1) Those pixels where there is a luminosity variation are selected.

　m.2) Interesting pixels are saved in an appropriate table while interesting and non interesting pixels are stored in an other table for secondary and complementary analysis [4].

**n) Second Level Trigger** (Selection of Microlensing Events).

n.1) Symmetry test. Except in the case of a multiple lens or star, the light curve should be symmetric in time around the maximal amplification.

　n.2) Achromaticity test. Gravitational lensing is an achromatic phenomenon. However, the lensed star has not, in general, the same colour as the background and only the luminosity increase is achromatic (in constant seeing condition):

$$\frac{\Delta F_{pixel}^{red}}{\Delta F_{pixel}^{blue}} = \frac{F_{star}^{red}}{F_{star}^{blue}} = \text{constant in time} \quad . \tag{15}$$

From a first approximation point of view we can distinguish between lensing events, novae and supernovae by the point (n.1), and between lensing events and variable stars by (n.2).

**o) Pilot Analysis** (Quasi-On-Line). At this point, the analysis must be performed immediately after the DAQ and by automatic procedure without any assistance of the operator.

o.1) Paczynski fit (single point-like lens, single point-like source)[5].

　o.2) Fit with a binary point-like source and single point-like source).

　　The best fit is obtained by making a $\chi^2$ minimization using the parametric lens model as reported by S. Mao e R. Di Stefano [12] or M. Dominik [13].

　o.3) Fit with a single point-like lens and binary point-like source.

　　According to Griest & Hu[14], in this fit, there will be two $t_0$ and two $u_0$ and so two functions: $u_1(t)$ and $u_2(t)$. Consequently, if we define:

---

[4] We note that we need to save all pixel information, since some pixels which are not interesting in an exposure could be good candidates in the following observations. Such pixels, in the previous observations, will be useful to evaluate $\phi_{BKG}^{ij}$. For these reasons, it is useful to divide the data inside the database in two tables connected each other with an appropriate relational link.

[5] A good method for making this fit is to follow the same procedure that AGAPE collaboration used for its data selection, [7].

$$\omega = \frac{L_2}{L_2 + L_1} \quad , \tag{16}$$

the amplification will be:

$$A_{binary}(u_1(t), u_2(t)) = (1-\omega) A_{single}(u_1) + \omega A_{single}(u_2). \tag{17}$$

In other words, the light curve will be the superposition of two light curves for two point-like sources behind the lens. If we denote $2\rho$ the distance between the closest approaches of the lens to the two components and $2\lambda$ the distance between the source components, both measured in projected Einstein radii $r'_E = \frac{d_{OS}}{d_{OL}} r_E$, it follows that

$$\rho = \frac{t_{0,2} - t_{0,1}}{2 t_E} \quad , \tag{18}$$

and the angle $\beta$ between the direction from source component 1 to source component 2 and the lens trajectory is given by:

$$\beta = \tan^{-1}\left(\frac{u_{0,2} \pm u_{0,1}}{2\rho}\right) = \tan^{-1}\left(\frac{u_{0,2} \pm u_{0,1}}{t_{0,2} - t_{0,1}} t_E\right). \tag{19}$$

At the end the half-distance between the components is

$$\lambda = \rho / \cos\beta \quad . \tag{20}$$

If we define $p(t) = \frac{t - t_0}{t_E}$ for a single lens, we have:

$$p(t) = \frac{1}{2}[p_1(t) + p_2(t)] \quad , \tag{21}$$

where

$$p_1(t) = \frac{t - t_{0,1}}{t_E} \quad \text{and} \quad p_2(t) = \frac{t - t_{0,2}}{t_E} \quad .$$

**p) Off - line Analysis.**

The surviving events after the first level trigger are interesting since there is an amplification in the light curve. About the fraction of these events that do not survive after the second level trigger we can say that they are non lensing events, except some kind of events that we will describe below. For such events, it is interesting to execute a specific and fine analysis for distinguishing among:

p.1) Events related with supernovae and novae;

p.2) Events related with a variable star;

p.3) Events related with a single point-like lens and an extended uniform bright source (constant density inside a finite volume $\rho(\mathbf{x}) = \cos t$);

p.4) Events related with a single point-like lens and an extended source $(\rho = \rho(\mathbf{x}))$;

p.5) Events related with a binary point-like lens and an extended uniform bright source ($\rho(\mathbf{x}) = \cos t$);

p.6) Events related with a binary point-like lens and an extended bright source $(\rho = \rho(\mathbf{x}))$;

p.7) Events due to an extended lens;

p.8) Events due to a multiple lenses system;

p.9) Detection of extrasolar planets.

At the moment, we believe that these events cannot be included in the point o) taking into account their peculiarity. However this approach, using a pre-selection (first level trigger) allows to know soon which events could be interesting for both studies: quasi-on-line and off-line. In this way there is no analysis on non interesting pixels with a huge time reduction. So even taking into account the first level trigger without the point o) the researcher obtains relevant advantages with respect to a traditional analysis where it need the study of all pixel for each observation.

About the point p.1) - p.9) some works are in progress to build an off-line analysis starting from the Analysis Unit shown in the previous scheme. However a specific paper will fix all these aspects [15].

From the description of this project, we understand the relevance of a strong and effective database: it should be able to manipulate the date very quickly. For this reason, probably, we shall use an object oriented database that has all the advantages of the object oriented environment. In fact, in this way, we shall build a database by following our specific request, orienting the data storage by taking into account the specific kind of analysis to do. However more details about this subject will be discuss in a forthcoming paper [16].

## 6 Conclusion and Perspective

The achievement of such an environmental project allows us to obtain:

- to take into account the geographic position, the new generation optics and device at the Toppo telescope, the first large microlensing survey performed in the North-Hemisphere for spiral arms observations and marginally for the bulge;
- to obtain a detailed survey on several galaxies besides the Galaxy (first of all M31);
- to reach the capability of detecting planets (both Jupiter-like and Earth-like);
- the partecipation in the follow-up observations of microlensing events which will be announced by the Global Microlensing Alert Network (GMAN) and Planet Collaboration;
- the possibility to use larger or spacecraft telescope as HST to resolve interesting pixels obtaining more astrophysical information on the amplified object.

The first steps for building this environment will be the described in [15], [16].

To conclude we stress the importance of an on-line selection and a quasi-on-line analysis; in fact it could give us interesting result soon after the observations and DAQ. It becomes indispensable if one wish to partecipate in an Alert program or if one wish to have some time on larger or spacecraft telescope.